# Balaton Borders: Data Ceramics for Ecological Reflection

Hajnal Gyeviki
Independent Artist, Hungary
gyeviki.hajnal@gmail.com

Mihály Minkó
Moholy-Nagy University of Art and Design, Hungary
minko.mihaly@mome.hu

Mary Karyda
Moholy-Nagy University of Art and Design, Hungary
maria.karyda@mome.hu

Damla Çay
Moholy-Nagy University of Art and Design, Hungary
damla.cay@mome.hu

## Abstract
Balaton Borders translates ecological data from Lake Balaton into ceramic tableware that represents human impact on the landscape, from reedbed reduction to shoreline modification and land erosion. Designed for performative dining, the pieces turn shared meals into multisensory encounters where food and data ceramics spark collective reflection on ecological disruption.

## Keywords
data physicalization, data edibilization, commensality, ecological design, environmental data, ceramic design, performative dining

## Introduction
Dining has long been a space not just for nourishment but for connection, dialogue, and decision-making. This project builds on the dining ritual to ask: How can commensality support engagement with urgent ecological data? Data often aim to inform or persuade, but in the context of environmental crisis, they must also foster conversation. Dialogue, after all, is the gateway to informed action. Lake Balaton, Hungary's largest freshwater lake, is increasingly threatened by human intervention: shoreline artificialisation, reedbed destruction, sediment accumulation, and unchecked urban expansion. In response to these pressures, this project uses data physicalisation to encode local environmental data into ceramic tableware and a performative dining experience. Each mug, jug, plate, and bowl expresses site-specific information through its material, shape, scale, and structure. Porcelain carries the data; food provides the context. Together, they compose a dining experience that invites interpretation through eating, handling, and conversation.

The contribution of this work is an immersive experience that bridges commensality – the practice of eating together [1] – and data physicalisation. By embedding ecological data into everyday rituals of dining, the project invites slower, more embodied engagement with environmental issues, rooted in community. Data ceramics invite reflection through the shared, sensory act of eating together, highlighting that what's on our plates is inseparable from the ecosystems we depend on.

collective care

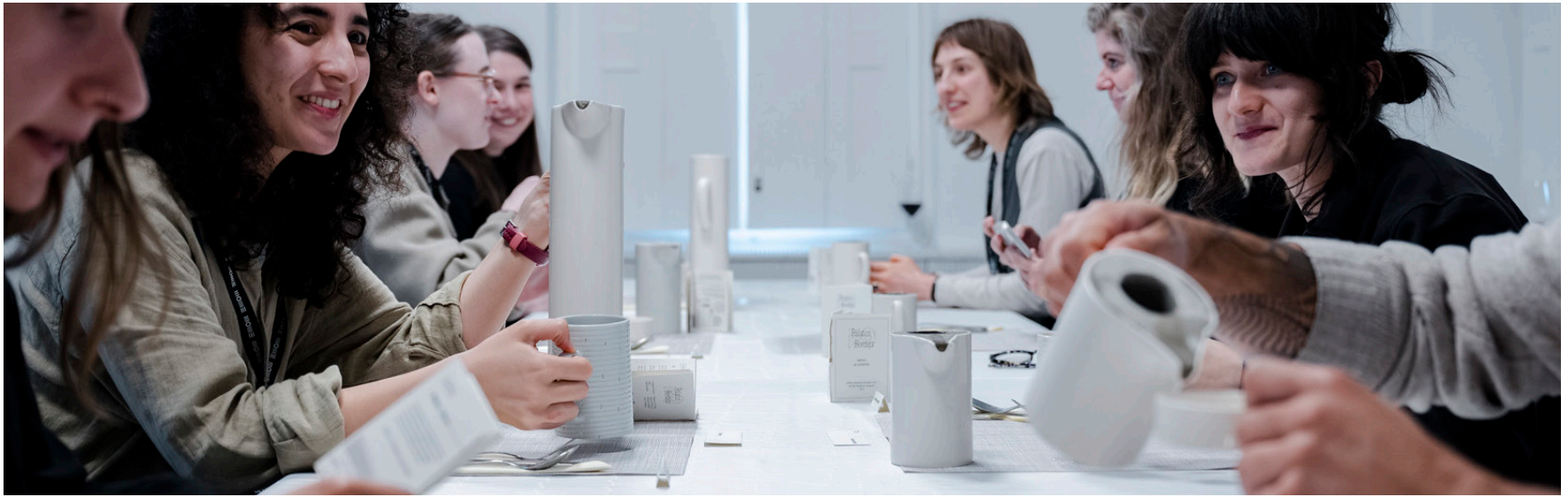

Fig. 1: Guests engage in lively conversation as they handle data-encoded mugs and jugs, turning the table into a space of shared ecological reflection. Credit: Máté Lakos

### Related Works

Data physicalisation, the practice of transforming data into tangible forms, has become an increasingly prominent approach in data communication. Since the term was introduced, it has been applied across a wide range of domains and contexts [2]. Several works have addressed ecological and critical issues through data physicalisation. Unknown Fields Division's Rare Earthenware [3] captures the hidden toxicity of electronic waste in the form of ceramic vessels made from mine tailings. Mischer Traxler's Access [4] visualises the unequal global distribution of drinking water by representing it through different sized glasses. These projects use everyday materials to communicate large-scale environmental problems in ways that invite critical attention.

The practice of data edibilisation builds on similar goals. Wang et al. [5] and Mueller et al. [6] describe how food can be used to represent data through taste, form, and materiality. Wilde and Karyda [7] argue that casual engagement with data through edibilisation can turn information into food for thought, exemplified by Li et al.'s Bitter Data [8], which mapped 100,000 distress postings from Chinese social media to the bitterness and color of tea. Other examples include Data Cuisine [9], where local ingredients visualise civic datasets, and Cyber Wagashi [10], which uses weather data to 3D print traditional Japanese sweets. These works create sensory, often social experiences that shift data interpretation into the realm of feeling, sharing, and speculation.

Ceramics also appear in the growing field of data craft. Nicenboim's [11] Form Follows Data mugs and Guljajeva et al.'s [12] Loading Ceramics project explore how personal and scientific data can be embedded into the material language of clay. Thudt et al. [13] describe data craft as a way of integrating data into everyday life and reflection.

Balaton Borders build on these works by combining ceramic data artefacts with performative dining, translating environmental data into a series of tactile, edible, and contextualised encounters. While previous work has explored food or craft individually as modes of data engagement, this project brings them together through commensality, treating the shared meal as a critical interface for ecological storytelling and collective reflection.

## Making of the The Tableware
*Context and Concept*

Balaton is Hungary's largest freshwater lake and a site of intense recreational, cultural, and economic activity. It's also one of the clearest examples of accelerated ecological degradation due to human intervention. The lake's reedbeds have been fragmented (See Fig. 2). Stone and concrete strips now line the shores, transforming the lake's natural edge into hardened infrastructure. Agricultural runoff from the surrounding slopes has intensified erosion and sedimentation, leading to nutrient overload and recurring algal blooms. In formerly flood-prone zones that once belonged to Ancient Lake Balaton, unchecked construction has overtaken wetlands that once buffered ecological fluctuations. Each of these interventions reflects a broader pattern: a rupture in the long-standing relationship between human activity and ecological rhythm. These are not isolated problems but symptoms of a mindset that treats the landscape as expendable.

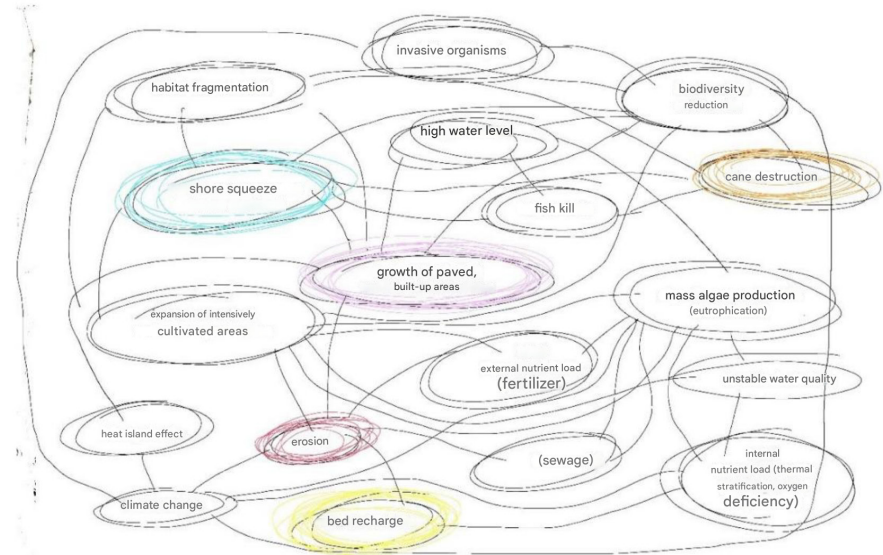

Fig. 3: First interpretation of the relationships between anthropogenic impacts on the ecology of Lake Balaton

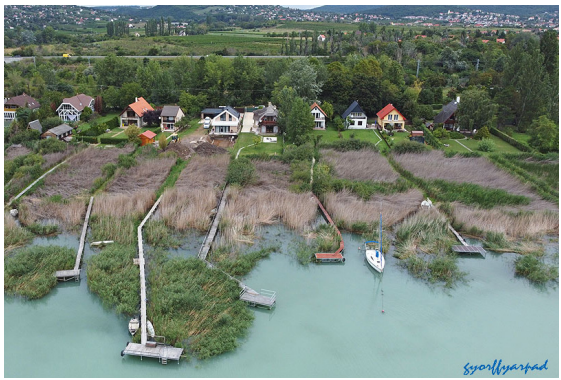

Fig. 2: Reedbeds around Lake Balaton are fragmented by private piers. Credit: Győrffy Árpád

The idea for the project took shape during a course called Ecological Data Visualization. The course coincided with research initiatives under the Veszprém-Balaton 2023 European Capital of Culture program, where students were asked to find data and propose ways to visualize it. A shared resource among the students was a scientific paper from Pomogyi [14] on direct human interventions in the coastal vegetation of Lake Balaton. Within it, the dataset that represented the number of reedbed cuttings per municipality was shocking for the first author who is also the artist who developed the data ceramics. The scale of ecological disruption, and that emotional response became key elements of the design process and outcome.

The first concept to emerge was a set of perforated mugs, each mapping the reedbed cuts as visible punctures spiraling around its surface. Since then the project expanded to have a full tableware including mugs, jugs, deep plates, flat plates and small plates, which encoded site-specific data through their shape, volume, and functionality or dysfunctionality. The decision to make the tableware dysfunctional came from a desire to resist smooth consumption and provoke a moment of disturbance. A jug might be too heavy to lift or too full of concrete to pour. A plate might be tilted just enough to spill. These moments of discomfort interrupt the flow of dining and make visible the invisible disruptions that ecological data represents. The project asks what happens when we bring the dysfunctions of our ecosystems to the table and invite others to sit with them.

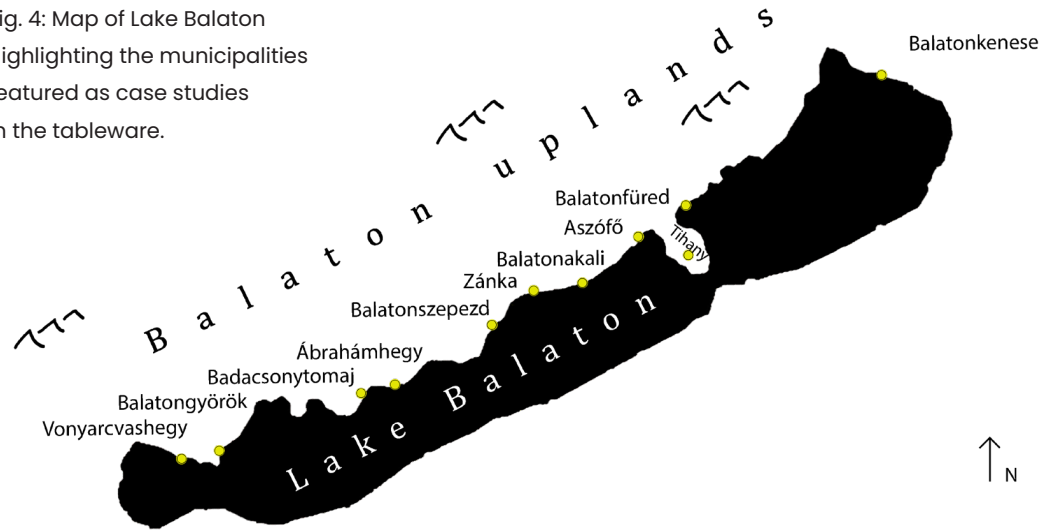

Fig. 4: Map of Lake Balaton highlighting the municipalities featured as case studies in the tableware.

To translate these findings into ceramic form, spatial and numerical data was visualized using QGIS (See Fig. 5). These values were then brought into Rhino 3D, where mugs, jugs and deep plates were modeled to reflect a specific parameter such as height, volume, and cut patterns. Flat plates, on the other hand, were visualized in Illustrator, as their design involved translating spatial data into angles. However, not all data sets allowed for seamless translation. Some lacked consistent resolution across municipalities. Others required reformatting or interpretation to align with the logic of physical form. Material and fabrication constraints also shaped final outcomes: the strength and balance of porcelain, tolerances in casting required careful adjustments. In many cases, values were scaled to maintain a balance between data integrity, clarity, and the physical limits of ceramic design.

*Data*

The selection of ecological phenomena was guided by both conceptual relevance and the availability of public, municipality-level datasets that could be visually and materially translated. Each process needed to reflect a pressing anthropogenic impact on Lake Balaton's ecosystem and be mappable in form, scale, or spatial pattern. From an initial survey of regional challenges, five key phenomena were chosen: reedbed cutting, shoreline artificialisation, erosion from agriculture, sedimentation, and built-up areas in the ancient lakebed. Each dataset came from a specific source with varying levels of granularity and consistency.

Reedbed and shoreline data were drawn from Pomogyi [14], which provided municipality-level figures on the extent of reedbeds and hardened shoreline segments. Slope erosion was assessed using Google Earth Pro's cross-section tool and Copernicus land use maps, allowing land gradients and vegetation cover to inform the shaping of deep plates. The Copernicus Global Human Settlement Layer (10-meter raster) was used to visualize built-up areas in the ancient lakebed, which directly shaped the design of the small plates. Aerial imagery from Google Maps supported the modeling of the coast lines in the flat plates.

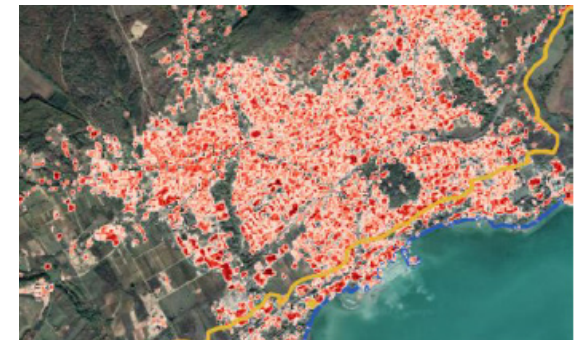

Fig. 5: Visualization from QGIS showing built-up areas in Balatonfüred. The zone between the yellow and blue lines marks the Ancient Lake Balaton.

*Design and Making*

The project began with the development of mugs, initially designed to visualize reedbed degradation using data from Pomogyi [14]. Each mug encoded three variables: total reedbed length, number of reedbed cuts, and the average distance between cuts per municipality (See Fig. 6). The total lengths of the reedbeds were scaled down at a 1:1000 ratio and translated into a spiral with a 0.5 cm thread pitch wrapped around the cylindrical body of each mug, all sharing the same diameter. The number of reedbed cuts was represented as perforations placed along this spiral. Mug height reflected total reedbed length, and the placement of perforations corresponded to calculated average distances. Because of the precision required, the mugs were modeled digitally in Rhino, 3D printed in PLA, and cast in porcelain using individual plaster moulds (See Fig. 8).

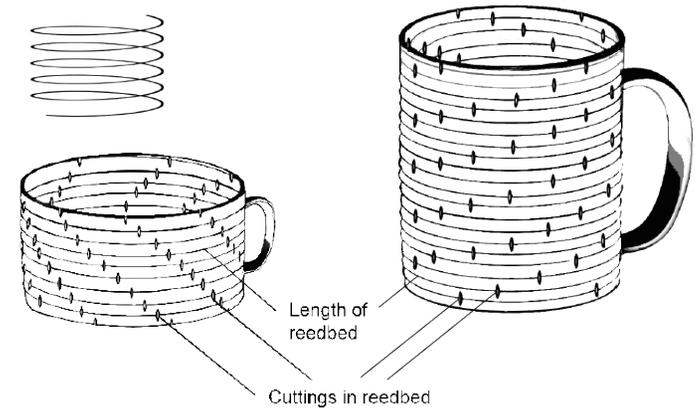

Fig. 6: Concept sketch of mugs.

Later, the jugs were developed to represent the full shoreline length of each municipality (See Fig.7). Their height was calculated proportionally to the mugs, based on the total coastline length. To express the extent of artificial shoreline, concrete was later poured into the jugs. This formed a visible cross-section from above: The proportion of concrete reflected the ratio of hardened shoreline to total shoreline. A jug with no concrete indicated a natural shoreline, while one filled entirely with concrete indicated complete artificialisation. The concept was expanded into a full tableware set. Three new pieces were added: small plates, deep plates, and flat plates. Each addressed a distinct ecological issue around Lake Balaton and aligned with a specific tableware function.

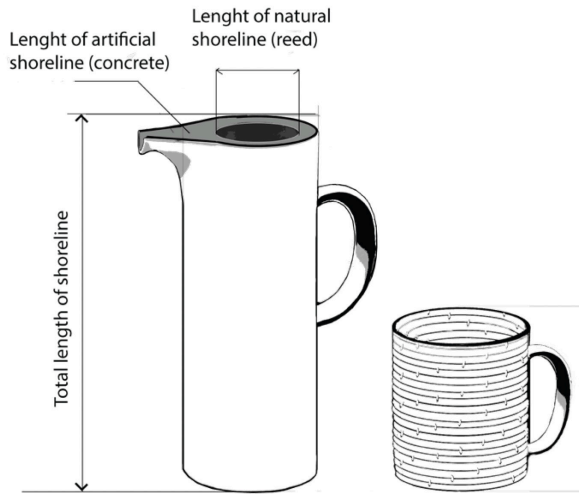

Fig. 7: Concept sketch of mugs and jugs.

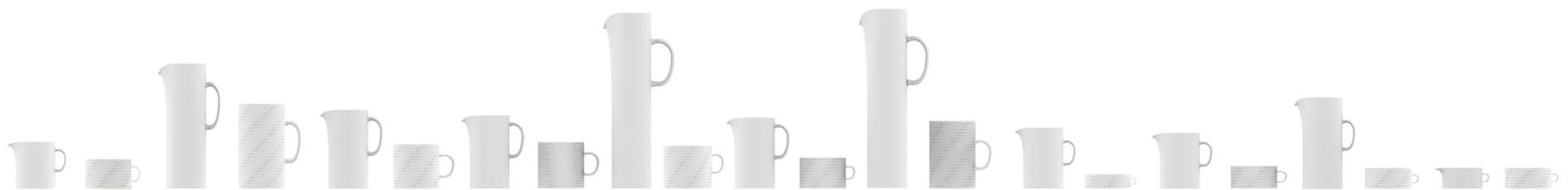

Fig. 8: 3D Renders of the mugs and jugs.

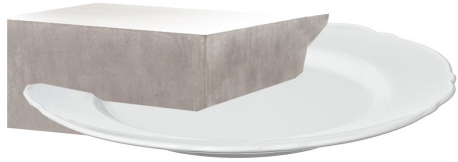

Fig. 9: 3D Render of the small plate.

The small plates visualized built-up areas within the ancient lakebed zone, a highly visible ecological and political concern in Hungary. The idea to pour concrete onto the plate (See Fig. 9) surface emerged from a desire to express the environmental impact of construction by mimicking the irreversible covering of ecosystems through real estate development. Concrete and ceramic became visual equivalents for built and unbuilt land, turning each plate into a pie chart of land-use transformation (See Fig. 12). Built-up surface data was sourced from the 2018 Copernicus Global Human Settlement Layer (10 m resolution raster).

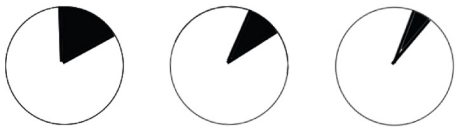

Fig. 10: Pie charts for the small plate.

Values ranging from 0 to 100 indicated the proportion of urban development within each 100 m² grid. These values were mapped onto the plates in the form of concrete segments, functioning visually as pie charts (See Fig. 10).

The frame of analysis was not defined by municipality borders but instead followed the elevation-defined outline of the Ancient Lake Balaton (113 m above Adriatic sea level), based on the guidance of ecological experts (See Fig. 11).

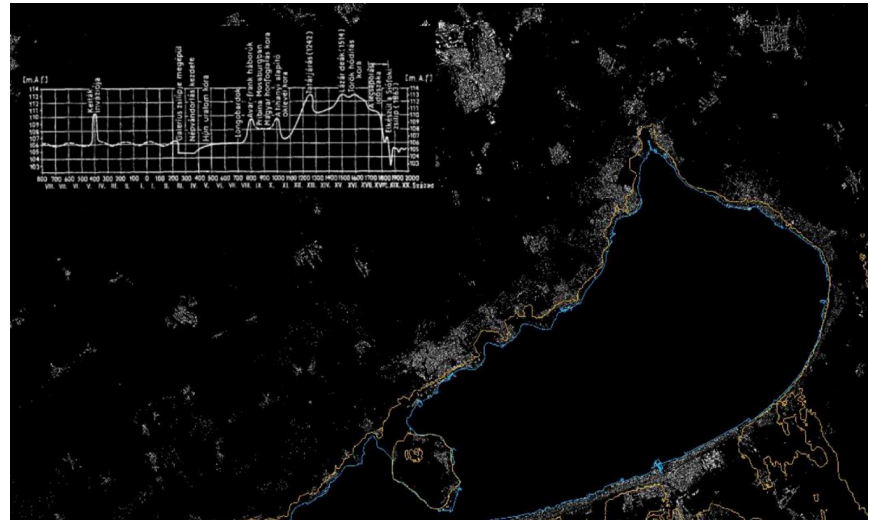

Fig. 11: Map exported from QGIS: the blue line shows the current shoreline of Lake Balaton, while the yellow line marks the extent of Ancient Lake Balaton, an area experts recommend for ecological restoration.

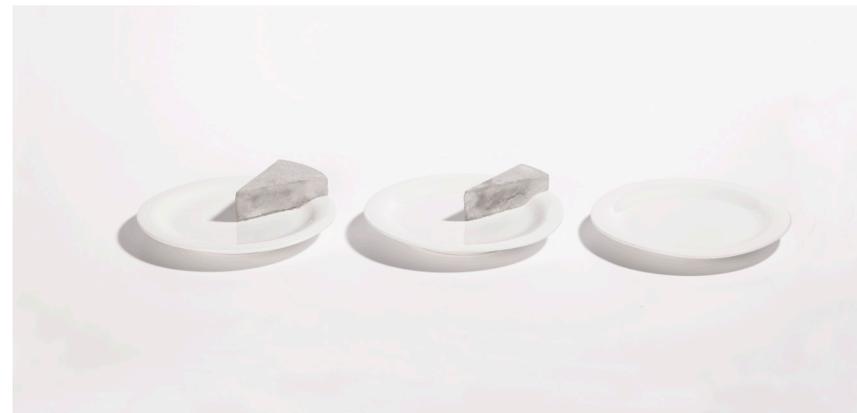

Fig. 12: Small plates from three municipalities visualizing built-up areas through concrete segments.

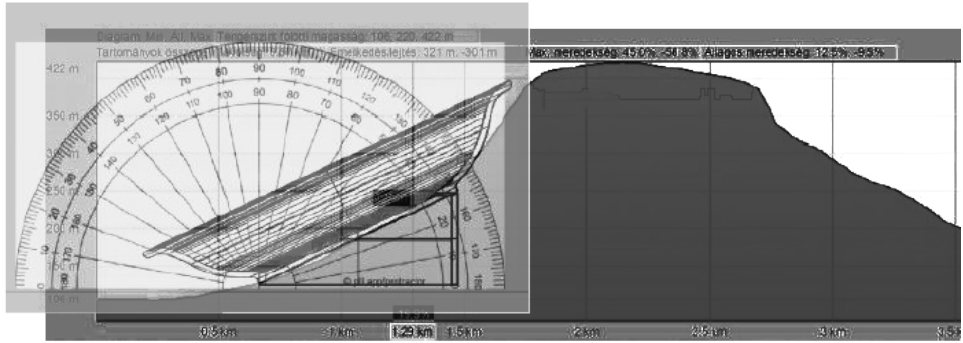

Fig. 13: Elevation profile of Balaton's northern shore extracted from Google Earth Pro, used to calculate slope angles for the deep plates.

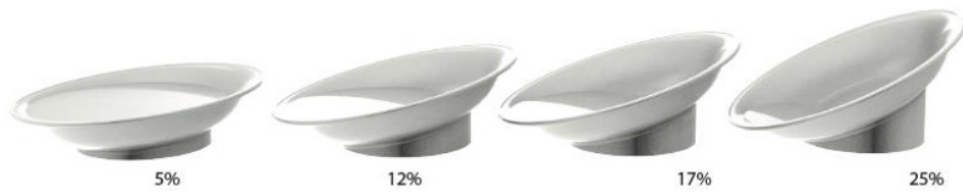

Fig. 14: Porcelain deep plates cast at varying angles (5%, 12%, 17%, 25%) to represent differing levels of erosion across municipalities.

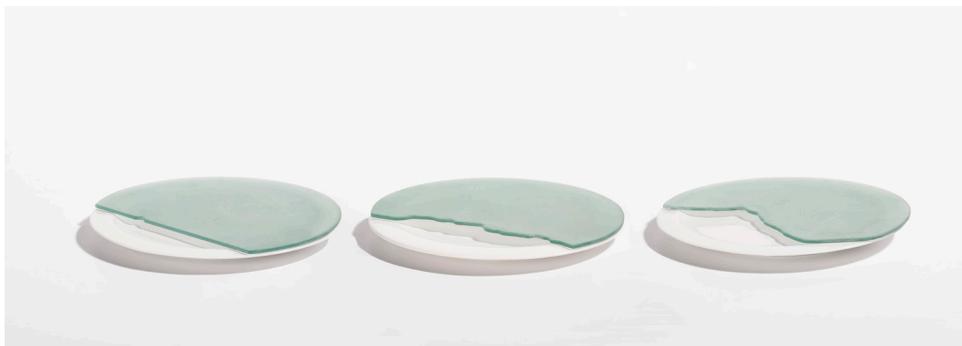

Fig. 15: Flat plates overlaid with blue-hue glass sheets shaped according to shoreline outlines, visualizing sedimentation and eutrophication patterns.

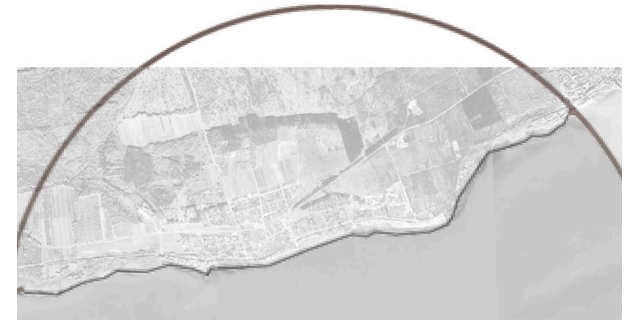

Fig. 16: A shoreline outline to be mapped on glass sheets.

The deep plates addressed erosion and soil degradation. Their sloping form was derived from elevation profiles exported from Google Earth Pro, taken perpendicularly to the lake (See Fig. 13). These profiles informed the incline of each plate, allowing the experience of reduced volume of food to serve as a metaphor for topsoil loss (See Fig. 14). The plates were produced from a single plaster mould and the cast porcelain plates were cut in varying angles to create different degrees of slope. The flat plates were designed to address sedimentation and eutrophication (See Fig. 15). A blue-hue glass sheet was added, resembling the distinct color of the lake and its coastal contour of the corresponding municipality, scaled at 1:22,400 (See Fig. 16). This layer referenced both the loss of topsoil from surrounding farmland and the spread of algal blooms. This addition aimed to reinforce the connection between land-based degradation and aquatic consequence, particularly sediment accumulation.

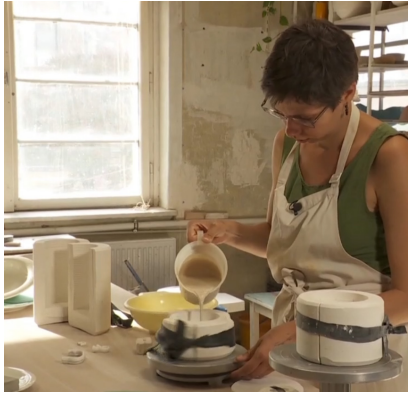 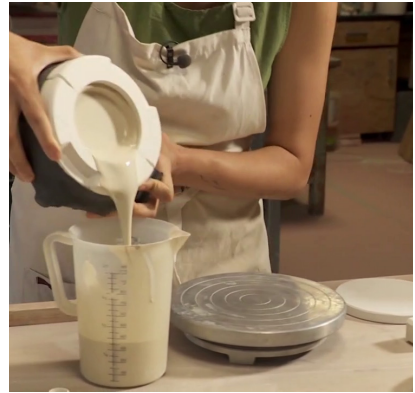 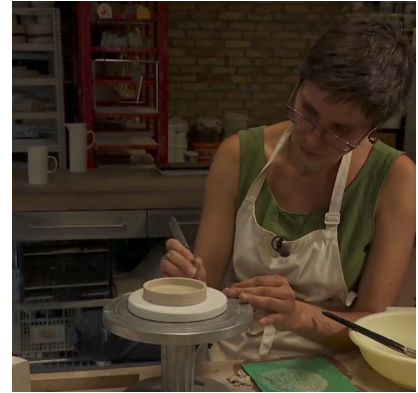 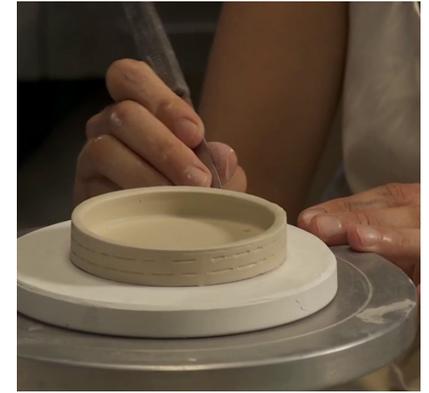

While the project critiques ecological degradation, its own material production involved trade-offs that raise questions about sustainability in practice. Each mug required its own custom mould. The jugs, however, used a more efficient idea: two base moulds (one tall, one short) from which volume could be adjusted by pour height. Handles and spouts were attached separately. Where possible, pre-existing forms were repurposed to reduce material waste. However, we acknowledge that the overall process relies on energy and resource-intensive methods such as ceramic firing, plaster moulding, and the use of non-renewable materials like PLA and concrete. Approximately 200 kWh of electricity was consumed in ceramic firing, and around 10 kilograms of cement were used in the jugs and small plates. These choices reflect a compromise between the conceptual goals of the work and the practical constraints of available tools, materials, and the artist's technical skillset.

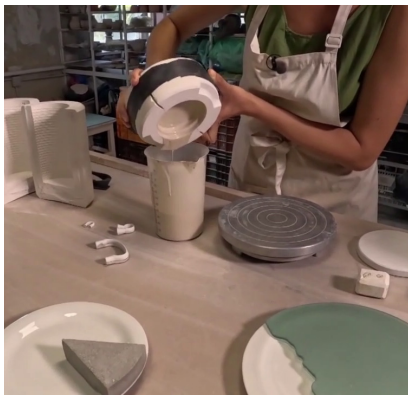 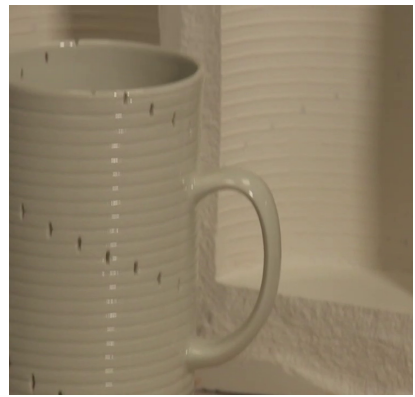 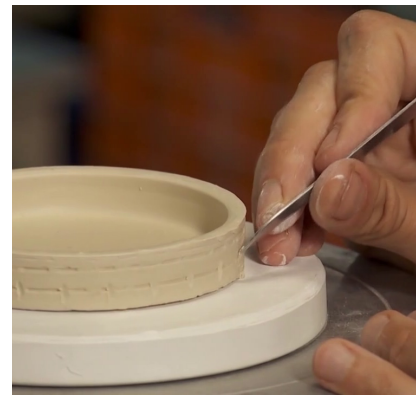 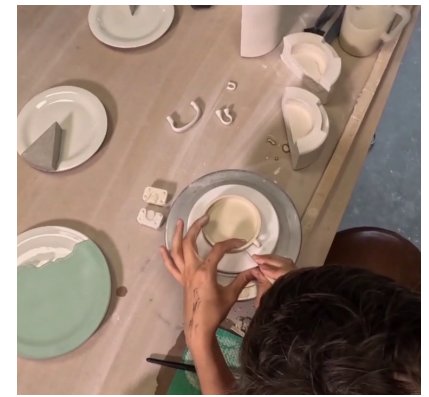

Fig. 17-24: Making process of the ceramic tableware: slip casting, mould preparation, finishing, and assembly of parts such as handles. Credit: M5 TV Channel, Péter Dömötör

## The Tableware

The current collection features tableware representing 11 municipalities along the northern shore of Lake Balaton: *Vonyarcvashegy, Balatongyörök, Badacsonytomaj, Ábrahámhegy, Balatonszepezd, Zánka, Balatonakali, Aszófő, Tihany, Balatonfüred*, and *Balatonkenese*. Each municipal set includes five pieces, with data mapped individually. In cases where original sources lacked coverage, such as *Aszófő* and *Balatonfüred*, data on reedbed cuts and shorelines were reconstructed using satellite imagery and QGIS software.

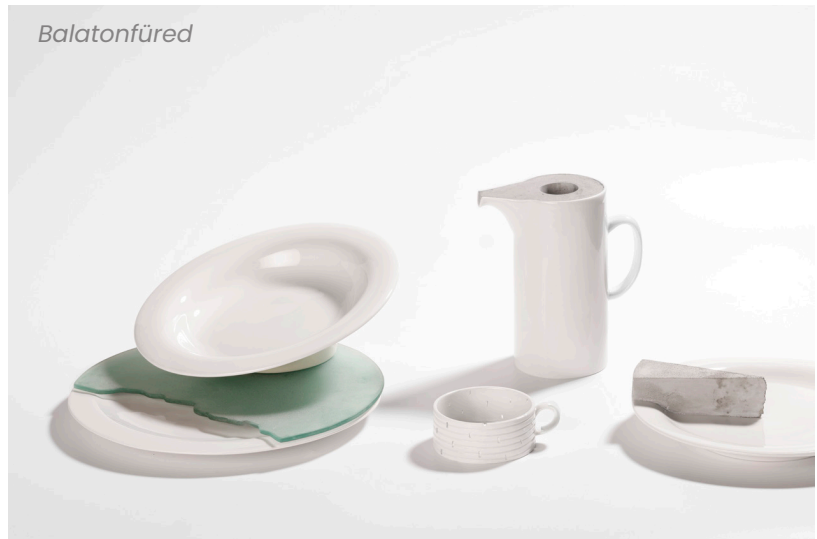

*Balatonfüred*

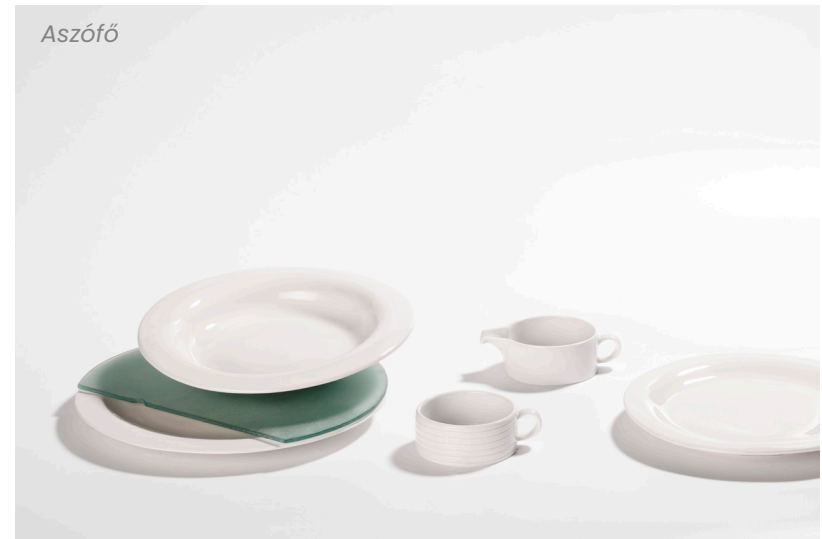

*Aszófő*

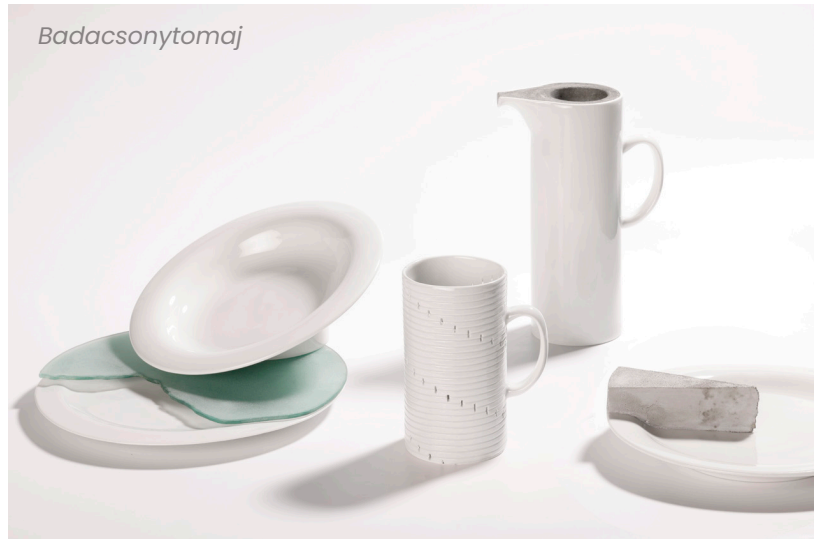

*Badacsonytomaj*

*Balatonfüred, Badacsonytomaj*, and *Aszófő* (See Fig. 25-27) present three distinct ecological profiles through their ceramic sets. *Badacsonytomaj's* deep plate has the steepest slope, indicating severe erosion, while *Balatonfüred's* plate is moderately tilted. Although the jug from *Badacsonytomaj* is taller, *Balatonfüred's* contains more concrete, reflecting its higher proportion of artificial coastline. The reedbed-to-coastline ratio in *Balatonfüred* is three times smaller than in *Badacsonytomaj*. *Aszófő*, in contrast, shows minimal ecological disruption. Its plate is nearly flat, the jug contains no concrete, and the mug has no perforations, representing the absence of shoreline construction and reedbed cuts. The built-up area is so small that it could not be meaningfully visualized on the plate, making *Aszófő* the least dysfunctional set in the collection.

Fig. 25-27: Three complete sets for Balatonfüred (top), Badacsonytomaj (bottom), and Aszófő (top right).

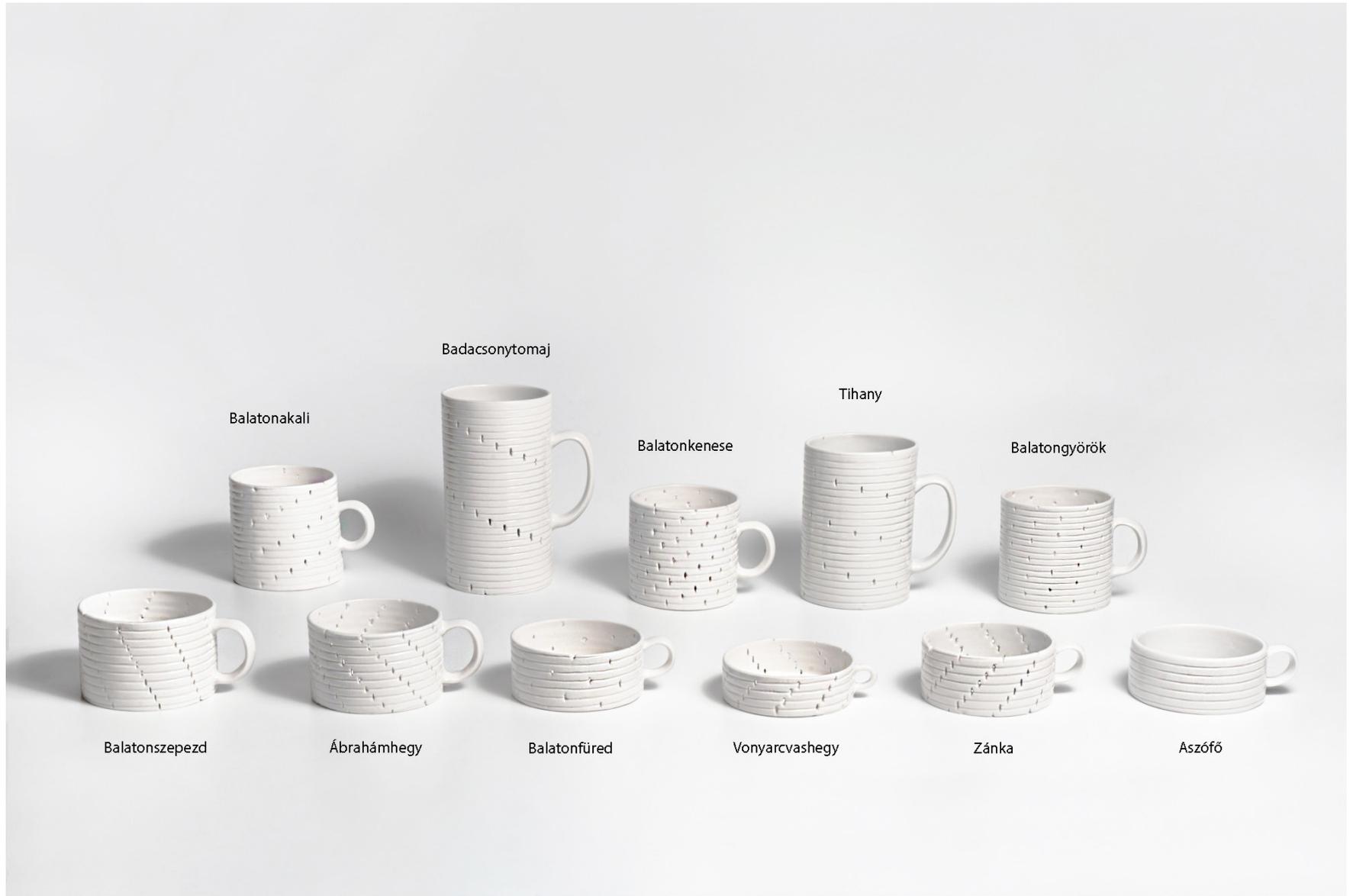

VISAP'25, Pictorials and annotated portfolios.

Fig. 28: Full set of mugs visualizing reedbed data from 11 municipalities: Balatonkenese has the highest number of cuts along the northern shore, Vonyarcvashegy has the shortest total reedbed length, and Aszófő shows no reedbed cuts.

## Dining Experience

The first performative dinner took place in December 2024 in Budapest and was co-organized with a philosopher who hosted the event. Eight invited experts attended, all with backgrounds in ecology or Balaton-related research. This was the first opportunity to test the tableware with uninitiated guests, and the event also served as the starting point for qualitative design research. Our question was: How these objects could function as conversation starters in different social contexts?

The dinner was a two-course vegan meal: pumpkin soup followed by chocolate cake. These choices were not optimal for highlighting the tableware's conceptual layers, but the focus remained on observing the guests' interaction with the objects.

One new piece was introduced specifically for this dinner: a large serving plate for the cake. Following the same pie chart logic as the small plates, it was intended to represent the average built-up surface across the Balaton region. However, since complete data for all 44 municipalities was not available, the average was calculated from the eight municipalities featured in the dinner. The result was a concrete segment just under one-eighth of the serving plate's surface, leaving one person with a visibly smaller slice. Intended as a provocative gesture, the moment was neutralized by the group when a guest who disliked sweets volunteered to take it.

The second Balaton Borders event took place in May 2025, as part of the EKSIG conference on data and visualization. Unlike the first, this event was open to all attendees, with 16 participants taking part in two sittings. This iteration emphasized food as a conceptual element. Mineral water from the Balaton Uplands was served (See Fig. 27), followed by vegan fish soup, halászlé (See Fig. 28), and the popular local flatbread, lángos. The use of lángos, which is round, flat, and tied to the region's culinary identity, reinforced the message of the small plates by presenting participants with only a portion of lángos, not a full one (See Fig. 29).

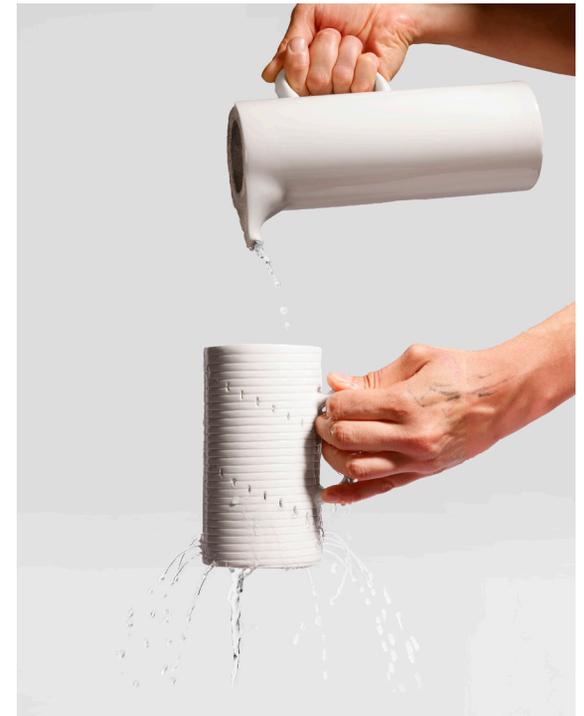

Fig. 27: First course: Mineral water from the Balaton Uplands, served in the jug and mug set

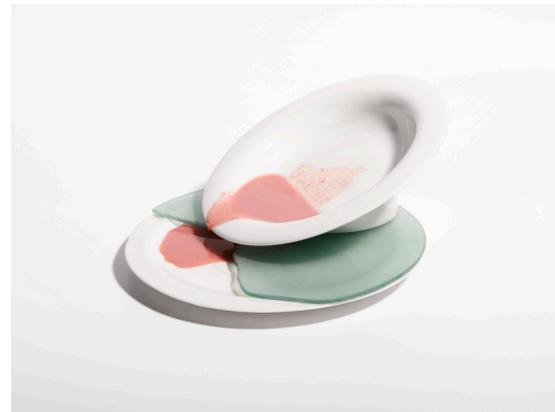

Fig. 28: Second course: Soup (halászlé), served in the sloped deep plate visualizing erosion

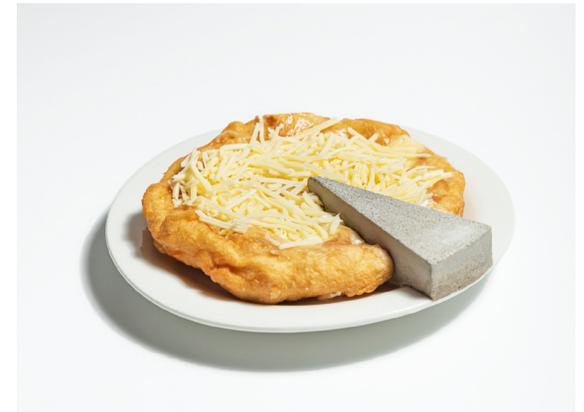

Fig. 29: Third course: Lángos, served on the flat plate visualizing built-up areas with a concrete segment

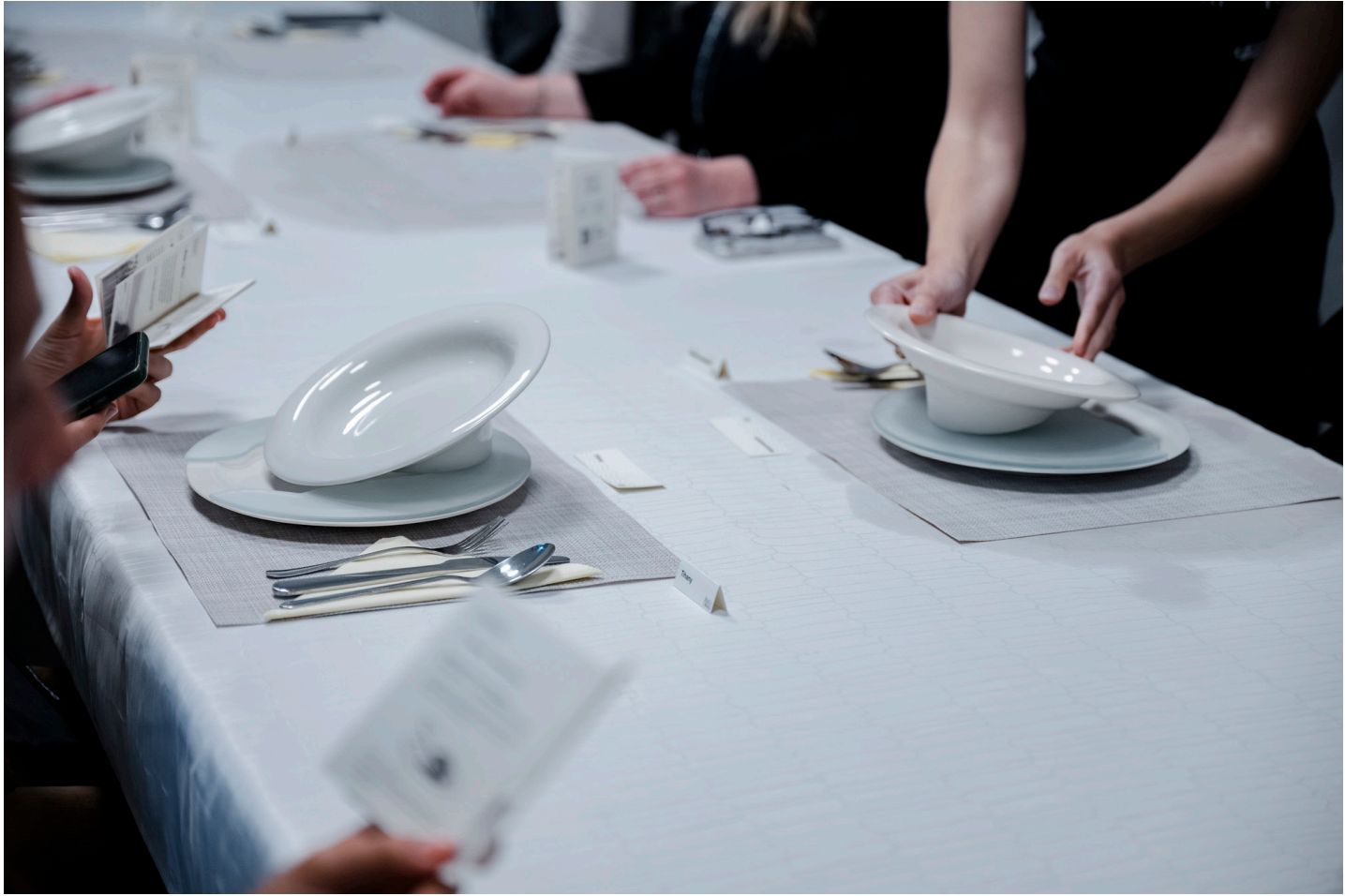
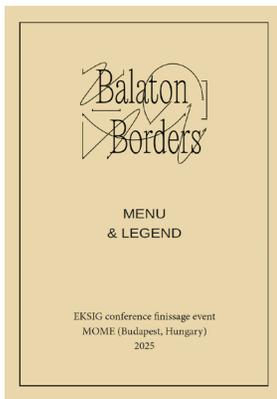
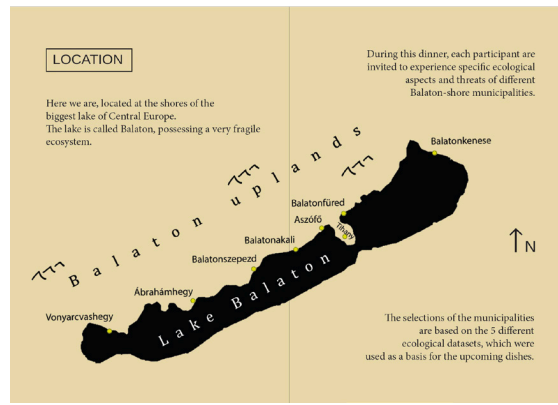
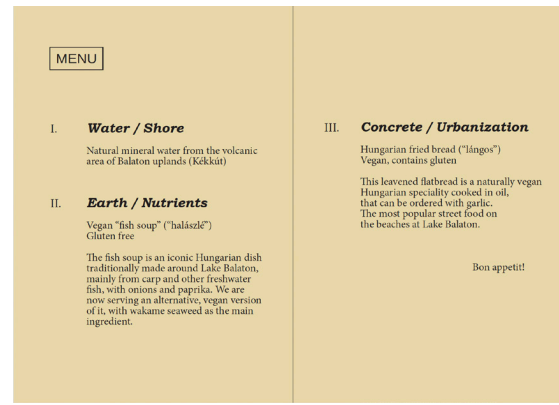

Fig. 30-33: Table setup before the dinner begins, featuring the explanatory booklet (left) that introduces the project and explains how each ceramic piece visualizes ecological data.

### Reflections on the Dinner Experience

The performative dinners marked a shift in the project's direction, from critique toward conversation. While earlier phases focused on materializing data through design, the dinners placed these objects in context, letting people physically and socially engage with the ecological realities they represent. This change deepened the project's intent, framing it not just as a visual or material commentary, but as an open-ended invitation to think, feel, and speak together.

Two dinner sessions were held during a conference in Budapest. The first group of participants volunteered by drawing seat cards at random, which created unexpected social pairings around the table. The meal began with a short welcome and contextual framing, introducing Lake Balaton as a fragile ecosystem and setting the tone for the experience. Participants were encouraged to explore the pieces without immediately reading the explanatory booklet. Many accepted the invitation to engage directly, lifting heavy jugs, attempting to drink from perforated mugs, or watching liquids spill across uneven surfaces. These early moments of hesitation and discovery sparked laughter, curiosity, and a range of emotional responses.

In the second round, guests entered differently. Rather than drawing cards, they simply found seats, many arriving early and spending time observing the space before the meal began. They appeared more comfortable using the tableware. They poured, spilled, experimented. Some tried eating soup from beneath the tilted plates. Almost everyone took photos or videos. The awkwardness of the objects became a shared point of interest, something to navigate together.

One interesting observation came from the participant who received a fully functional set, a jug without holes, an almost level soup plate, and an uncovered flat plate. She appeared visibly uncomfortable. While others struggled with the tableware, she could drink, eat, and serve without issue. She apologized to those seated nearby, sensing a disparity in experience. This unexpected moment highlighted how the presence of functionality, rather than its absence, can generate tension when seen against a backdrop of disruption. It also raised a design question that had been discussed in earlier phases: whether fully intact sets should be included at all. Their presence, as it turned out, became a quiet but powerful contrast.

What emerged across both dinners was not only a tactile engagement with data, but a space for reflection, discomfort, humor, and exchange. The objects worked as intended, not by delivering answers, but by holding questions. They prompted participants to think about land use, environmental loss, and responsibility, not in the abstract, but through the simple acts of eating and drinking. The table became a stage for shared inquiry, and the conversation became part of the design.

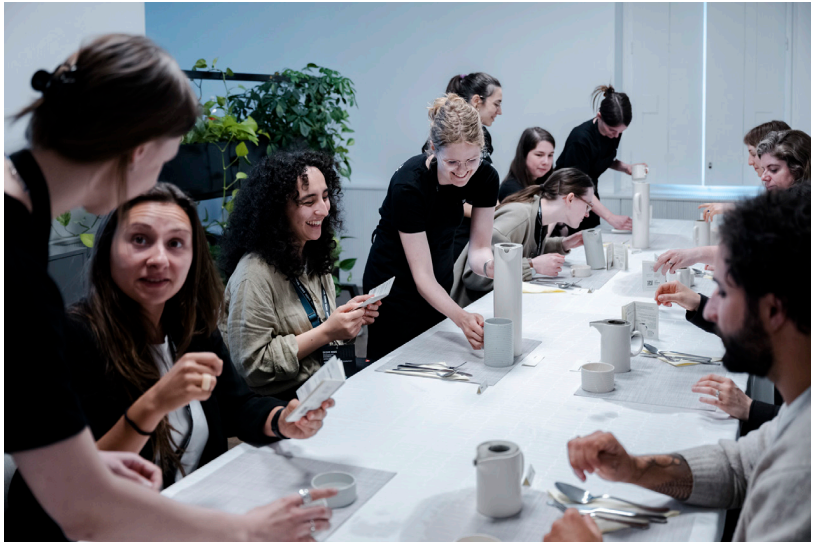

Fig. 34: Participants gather for the Balaton Borders dinner, engaging with the tableware and printed materials before the meal begins.

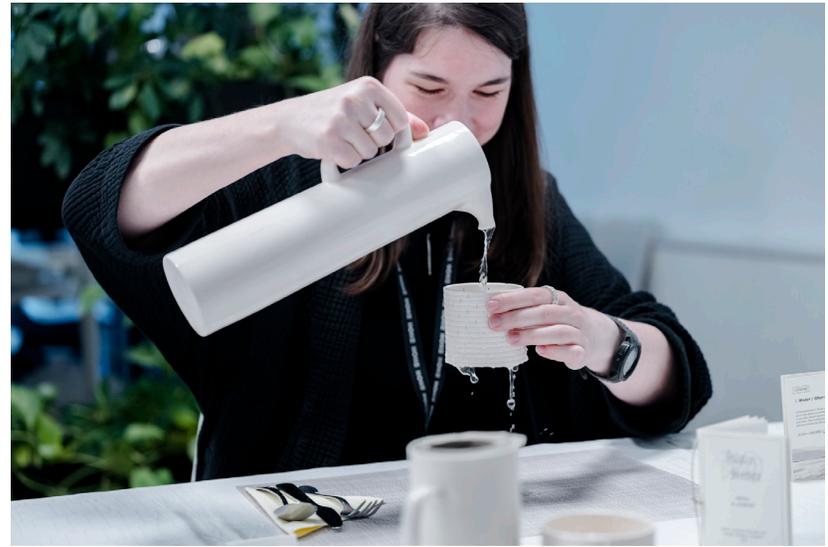

Fig. 35: A guest pours water from a perforated jug, causing it to spill onto the table, a disruption that reflects the ecological damage caused by shoreline modification.

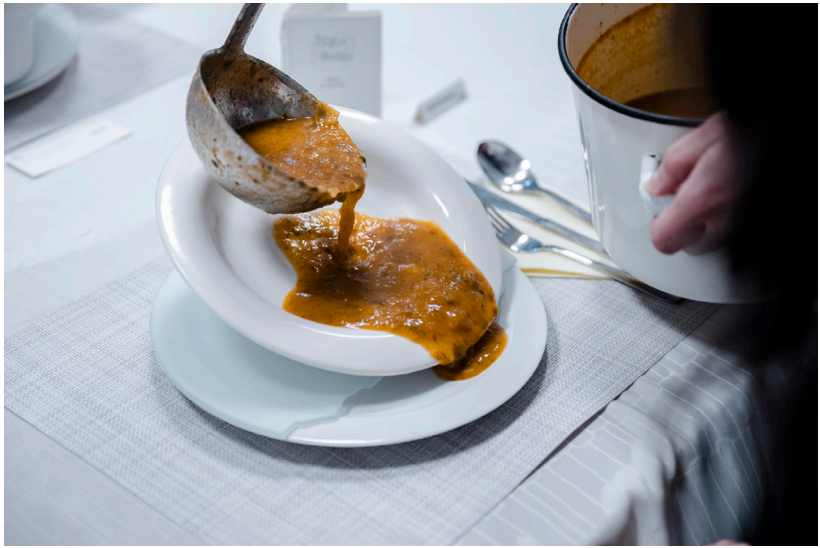

Fig. 36: Vegan fish soup being served into sloped deep plates, prompting reduced food volume and unstable presentation as a metaphor for topsoil loss.

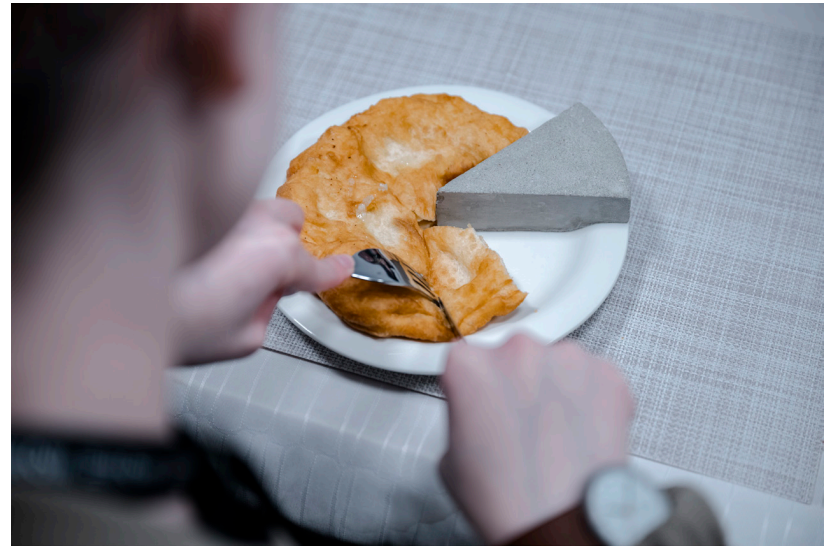

Fig. 37: Lángos served on a small plate partially covered in concrete, representing built-up land and limiting the portion to reflect the habitat loss.

## Discussion

This project demonstrates how data physicalization and edibilization can be integrated into a single, multisensory design language grounded in commensality. While previous works have explored these methods independently, either by translating data into physical artefacts or edible forms, this project brings them together around a central social ritual: the shared meal. Tableware and food come together to create a tactile and embodied interface with ecological data, where information is not only seen but also felt, handled, consumed, and discussed.

Through the performative dinners, the project shifted from a primarily critical stance to a more discursive one. Rather than merely confronting audiences with the consequences of ecological degradation, it sought to open space for dialogue and emotional reflection. The dinner table became a site for both discomfort and connection, where the struggle to eat or drink from dysfunctional vessels became a shared experience, rich with symbolism. The awkwardness of the tableware was not an obstacle but a deliberate provocation, designed to surface emotional and intellectual responses to complex ecological data.

In this setting, commensality functioned as both medium and message.

Participants engaged not just with the information embedded in each object but with one another, responding collectively to the unevenness, imbalance, and obstruction designed into the pieces. These responses of spilled water, frustrated attempts at pouring, laughter, quiet apologies, revealed the power of shared experience to humanize data and create lasting impressions. A participant's discomfort at receiving a fully functional set illustrated how absence of disruption, in a context defined by dysfunction, can itself become meaningful. Such moments underscored the nuanced ways in which material experience shapes perception and conversation.

This approach suggests a broader potential for performative data design. When ecological data is embedded into everyday rituals, particularly those rooted in community, it can bypass information fatigue and create deeper forms of awareness. The goal is not persuasion through facts alone, but engagement through presence. The objects do not dictate what to think; instead, they create conditions for thinking together. In doing so, the project contributes a method for activating ecological dialogue through design, one that is slow, embodied, and collective.

Future work may explore how such practices can scale or be adapted to other contexts. Equally important is the reflection on sustainability within the making process itself. The project acknowledges the contradictions of producing eco-critical work through resource-intensive methods and recognizes this tension as part of its discourse. Moving forward, questions remain about how data-based artefacts can be made more environmentally sound, without compromising their affective or narrative strength.

Ultimately, this project reframes the dining table as an interface for ecological storytelling. It invites others to sit down, to feel, to eat, and to think together.

## Conclusion

Balaton Borders shows how ecological data can move beyond charts and maps to become part of shared, sensory experiences. By translating environmental issues into physical and edible forms, we aimed to invite deeper reflection through the act of dining together. We hope the project inspires more explorations on how to turn data into conversation, and commensality into a space for collective reflection.

## Acknowledgements

This project was made possible through the support of many generous collaborators. Special thanks to András Mohácsi (supervisor of the diploma project), Mihály Minkó (mentor of the project and consultant of the diploma project), Júlia Vesmás (incubation programme lead), Zsuzsa Bokor, Mónika Csák, László Márhoffer, Sára Szeredi, Emese Mráz, Benjamin Balla A., Ferenc Kovács-Nagy (professionals in ceramic, plaster mould making, glass design and 3D printing), Mary Karyda, Damla Çay, Attila Bátorfy (visualization experts), Tamás Vig (QGIS and geospatial mapping expert), Ábel Szalontai (VEB2023 and Balatorium programme leader), Diána Berecz (field anthropologist expert), Anna Zilahi, Olga Kocsi (fine artist consultants), Piroska Pomogyi, András Zlinszky, Viktor Tóth, Ferenc Jordán (researchers), Balázs Fromm, Richard Usher, Szelina Bodócs, Máté Lákos (photographers), Máté Tóth-Heyn, Zsuzsanna Simon, Benedek Bognár (camera & video editors), Balázs Erlauer (co-director), Géza Kulcsár, Márk Horváth (theoretical). Finally, sincere thanks to all participants of the Balaton Borders events.